\documentclass[aps, twocolumn, showpacs]{revtex4}
\usepackage[english]{babel}
\usepackage{graphicx}
\usepackage{dcolumn}
\usepackage{color}

\def\beq#1{\begin{equation}\label{#1}}
\def\eeq{\end{equation}}
\begin{document}
\title{High-order optical processes: towards nonperturbative nonlinear optics}
\author{ V.V. Strelkov
 }
\affiliation{A. M. Prokhorov General Physics Institute of the RAS, Moscow 119991,  Russia \\
Moscow Institute of Physics and Technology (State University), 141700 Dolgoprudny, Moscow Region, Russia }
\begin{abstract}\noindent 
We develop an approach describing nonlinear-optical processes in the strong-field domain characterized by the nonperturbative field-with-matter interaction. The polarization of an isolated atom in the external field calculated via the numerical solution of the time-dependent Schr\"odinger equation agrees with our analytical findings. For the practically important case of one strong laser field and several weaker fields we derive and analytically solve propagation equations describing high-order (HO) wave-mixing, HO parametric amplification and HO stimulated scattering. These processes provide a way of efficient coherent XUV generation. Some properties of HO processes are new in nonlinear optics: essentially complex values of the coefficients in the propagation equations, the super-exponential (hyperbolic) growing solutions, etc. Finally, we suggest conditions for the practical realization of these processes and discuss published numerical and experimental results where such processes could have been observed.
\end{abstract}
\pacs{
  42.65.Ky            
  32.80.Rm            
}
\maketitle
\noindent

The nonlinear optics usually deals with the laser field which perturbatively interacts with matter. In this case the matter response is described with nonlinear susceptibilities $\chi^{(m)}$. This approach allows investigating numerous nonlinear optical effects involving few photons~\cite{Shen}. However, in the case of intense laser field the perturbation approach fails. A number of multiphoton process involving {\it electronic} dynamics (ionization, electronic rescattering, etc) are successfully described within nonperturbative approaches (like Keldysh approximation~\cite{Keldysh}), but the only fully {\it optical} process which is well-understood in this case is high-order harmonic generation (HHG). The study of the other nonlinear optical processes in the nonperturbative regime is limited, among other factors, by poorly developed theoretical methods of their description. For instance experimental~\cite{Spielmann_2010}  observation of the exponential growth of the high harmonic signal has led to an active discussion~\cite{Kazamias_2010, Kazamias_2011, Spielmann_2010_reply}.  The feasibility of high-order optical processes was already discussed in the early studies~\cite{Shkolnikov_1996, Chichkov_1999,Chichkov_2000, Eichmann_1995} but the theoretical methods were based on using nonlinear susceptibilities and thus were limited to processes of relatively low order. More recently the XUV amplification was obtained in simulations~\cite{Serrat_2013} based on direct numerical integration of the propagation equation with the nonlinear polarization calculated via numerical solution of the time-dependent Schr\"odinger equation (TDSE). 

In this paper we suggest an approach to describe nonlinear optical processes in the presence of a given strong laser field denoted as $E_0$. Processes involving other fields (denoted as $E_1$) are described with the susceptibilities, nonperturbatively induced by the pump field. Irrespective of the nature of nonlinearity, the microscopic polarization $P(t)$ is a {\it functional} of the external field: $P(t)=\Phi[E]=\Phi[E_0+E_1]$. The functional can be expanded in the Taylor series~\cite{Volterra} (applications in fundamentals of quantum mechanics~\cite{Feynman} and density functional theory~\cite{Engel} are well-known) using so called {\it functional} derivatives of the functional $\Phi$ over the function $E$. This expansion includes integration and thus differs from the traditional Taylor expansion of a function. The details of the calculation are presented in Appendix I. Considering the fields 
$
E_{0,1}=\mathcal{E}_{0,1} \exp(-i \omega_{0,1} t+\mathbf{k}_{0,1}\mathbf{r}+\varphi_{0,1})+ c.c.
$
with the real amplitudes $\mathcal{E}_{0,1}$, we expand the functional derivatives, as well as the functional $\Phi[E_0]$ in the Fourier series. Finally we find that 
\begin{equation}
\begin{array}{l}
P(t)=\sum_{q=-\infty}^{+\infty} \sum_{m=-\infty}^{+\infty} \tilde \kappa_{q}^{(m)} (\mathcal{E}_1)^{|m|}  \times \\
\exp\{iq(-\omega_0 t+\mathbf{k}_{0}\mathbf{r}+\varphi_0)+im(-\omega_1 t+\mathbf{k}_{1}\mathbf{r}+\varphi_1) \}  + c.c. 
\end{array}
\label{P_fin}
\end{equation}
where the induced susceptibilities are
\begin{equation}
\tilde \kappa_{q}^{(m)} (\mathcal{E}_0,\mathcal{E}_1,\omega_0,\omega_1) = \sum_{j=0,2,\cdots}^{+\infty} \kappa_{q}^{(m,j)} (\mathcal{E}_0,\omega_0,\omega_1)|\mathcal{E}_1|^j  
\label{tilde_kappa}
\end{equation}


Thus, the microscopic response is a sum of waves to which $q$ photons from the one field and $|m|+j$ photons from the other one contribute. For the case $\omega_0=\omega_1$, $\mathbf{k}_{0} \ne \mathbf{k}_{1}$ similar equation was found in~\cite{Platonenko_non-collinear}. Note that up to here we never supposed that the field $\mathcal{E}_1$ is small. Terms with $j=2,4,...$ in equation~(\ref{tilde_kappa}) can be understood as those describing processes where one, two, etc. photons from the field $E_1$  were absorbed and the same number of photons of this field were emitted.


When $\mathcal{E}_1<<\mathcal{E}_0$ the first term in equation~(\ref{tilde_kappa}) dominates, so we have 
$
\tilde \kappa_{q}^{(m)} (\mathcal{E}_0,\mathcal{E}_1,\omega_0,\omega_1) \approx \kappa_{q}^{(m,0)} (\mathcal{E}_0,\omega_0,\omega_1) 
$
As $\kappa_{q}^{(m,0)}$ does not depend on $\mathcal{E}_1$,  in the expansion~(\ref{P_fin}) the dependence on the field $\mathcal{E}_1$ remains {\it only} in the term $(\mathcal{E}_1)^{|m|}$. This makes this expansion for the polarization similar to the one appearing in the usual ("perturbative") nonlinear optics~\cite{Shen}. Below for brevity we write $\kappa_{q}^{(m)} \equiv \kappa_{q}^{(m,0)}$.



The remarkable practical importance of equation~(\ref{P_fin}) is that even in the strong field domain one can write and sometimes solve analytically (see below) the propagation equation with the right side given with equation~(\ref{P_fin}). So there is no need to calculate the non-perturbative response simultaneously with the propagation equation solution. Instead one should find the polarization using an appropriate theoretical approach (this can be numerical TDSE solution, strong-field approximation 
 for an isolated atom response, PIC simulations for plasma, etc),  calculate its spectrum $P(\omega)$ and then find $\kappa$ from this spectrum using equation~(\ref{P_fin}); alternatively, one can directly calculate $\kappa$ from equations~(\ref{kappa1})-(\ref{kappa3}) (see Appendix I).
\begin{figure}  
\centering
\includegraphics [width=1.0\columnwidth] {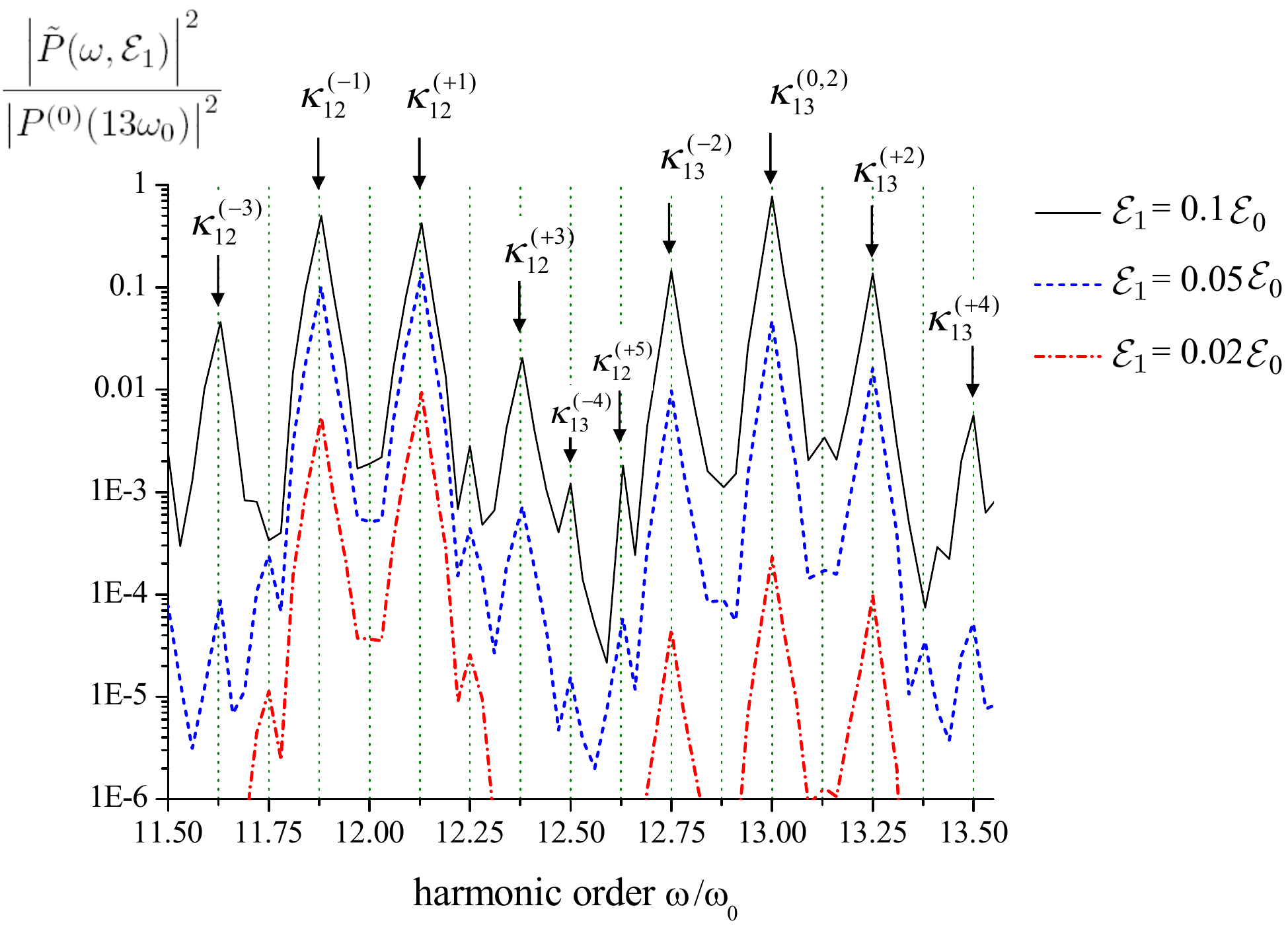}
\caption{(color online) Numerical spectra of the the polarization induced by the weak field $E_1$ in presence of the intense field $E_0$ for different strengths of the weak field, see text for more details. Arrows show the spectral components generated due to different induced susceptibilities $\kappa_q^{(m)}$. The spectra are normalized with the intensity of the 13-th harmonic generated in absence of the weak field.}
\label{Fig1}
\end{figure}

To check our analytical findings we compare the found properties of the polarization with numerical results. We solve numerically the 3D TDSE in the single-active electron approximation for a model Xe atom in an external laser field, the details of the TDSE solution are presented in~\cite{JPB_num}. For the given  strong field $E_0$ we calculate numerically the polarization $P$ in the presence of the weak field  $E_1$ and the polarization $P^{(0)}$ in the absence of this field. Figure~\ref{Fig1} presents the spectrum $ \tilde P (\omega , \mathcal{E}_1)=P(\omega, \mathcal{E}_1)-P^{(0)}(\omega)$ for the given $\mathcal{E}_0$ and different $\mathcal{E}_1$. The strong field wavelength is 800 nm and its intensity is $ 5 \times 10^{13}$ W/cm$^2$. The weak field has low frequency $\omega_1 =\frac{1}{8} \omega_0$ and low intensity, which differs in different calculations. 
\begin{figure}  
\centering
\includegraphics [width=0.9\columnwidth] {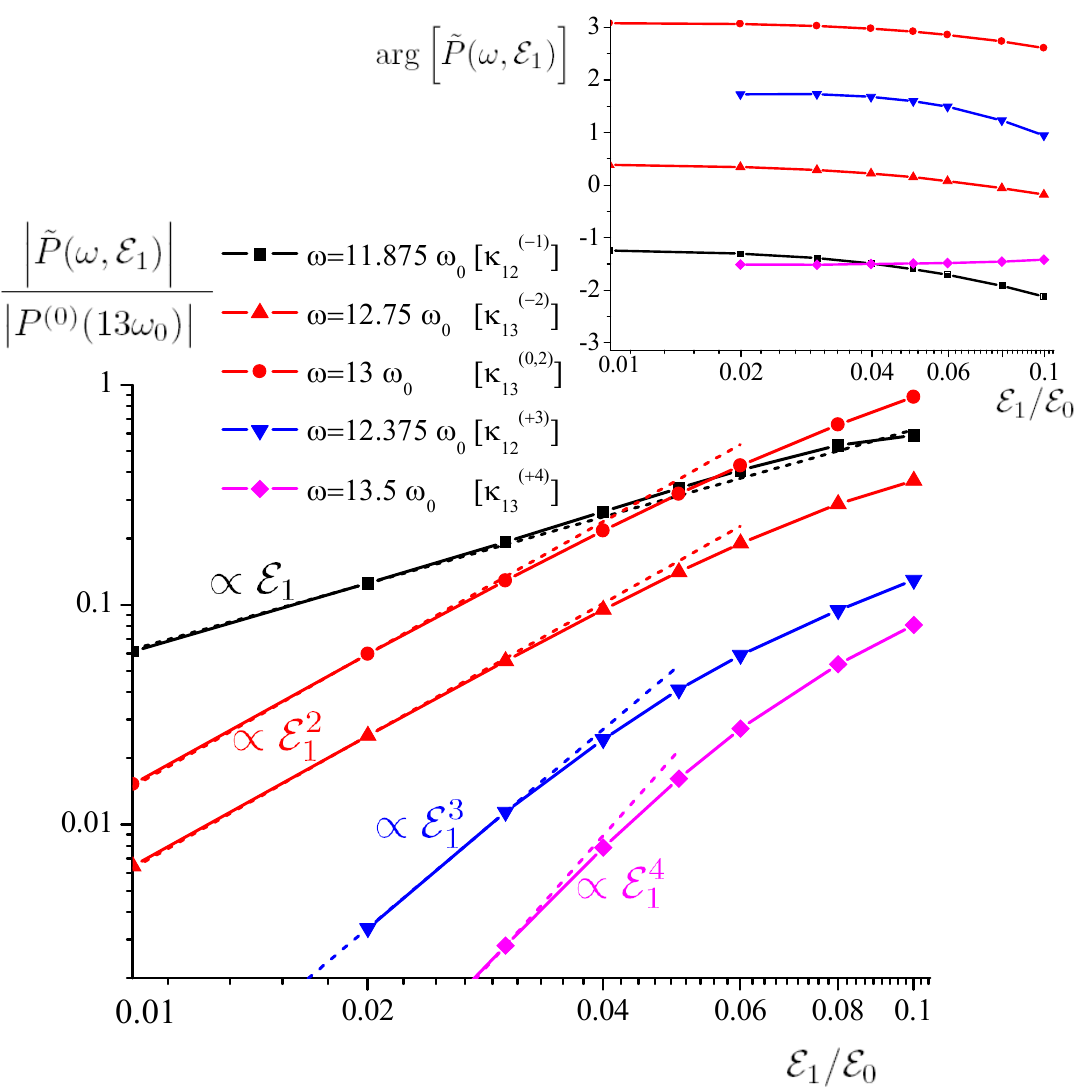}
\caption{(color online) 
Absolute values and phases of spectral components of the response shown in Fig~\ref{Fig1} as a function of the weak field strength. Several components  with different frequencies are shown, the induced susceptibility responsible for the generation of every component is presented in square brackets. Dotted lines show the power approximations.}
\label{Fig2}
\end{figure}

We can see that in agreement with the selection rule for a central-inverse medium, only processes with an odd number of photons $q+|m|+j$ contribute to the process (for $|m|\le 2$ and $j=0$ this selection rule was demonstrated already in the early experiment~\cite{Eichmann}). The strongest contributions are the "doublet" around the even harmonic of the pump field $q=12$ due to the $|m|=1$, $j=0$ processes 
, and the "triplet" around the odd one $q=13$ due to the $|m|+j=2$ processes.

Figure~\ref{Fig2} shows the absolute value and the phase of the response $ \tilde P (\omega , \mathcal{E}_1)$ as a function of the field $\mathcal{E}_1$. We can see that in agreement with our analytical findings the arguments are approximately constant and the absolute values increase approximately as $\mathcal{E}_1^{|m|+j}$, with lowest possible $j$. However, the deviation from the approximations increase with $\mathcal{E}_1$ because terms with higher $j$ should be taken into account. The role of terms with $j\ne0$ can be seen in this figure from the component with $\omega=13 \omega_0$ (red line with circles) generated due to $\kappa^{(0,2)}_{13}$. For $\mathcal{E}_1=0.1 \mathcal{E}_0$ we can see that $|\tilde P(13 \omega_0)| \approx  |P^{(0)}(13 \omega_0) |$, so  this contribution is not negligible for such $\mathcal{E}_1$. Below we shall deal with lower fields $\mathcal{E}_1$, thus studying only the processes with $j=0$.


In Appendix II we derive some properties of the induced susceptibilities, in particular, their permutation symmetry.

\begin{figure}  
\centering
\includegraphics [width=0.7\columnwidth] {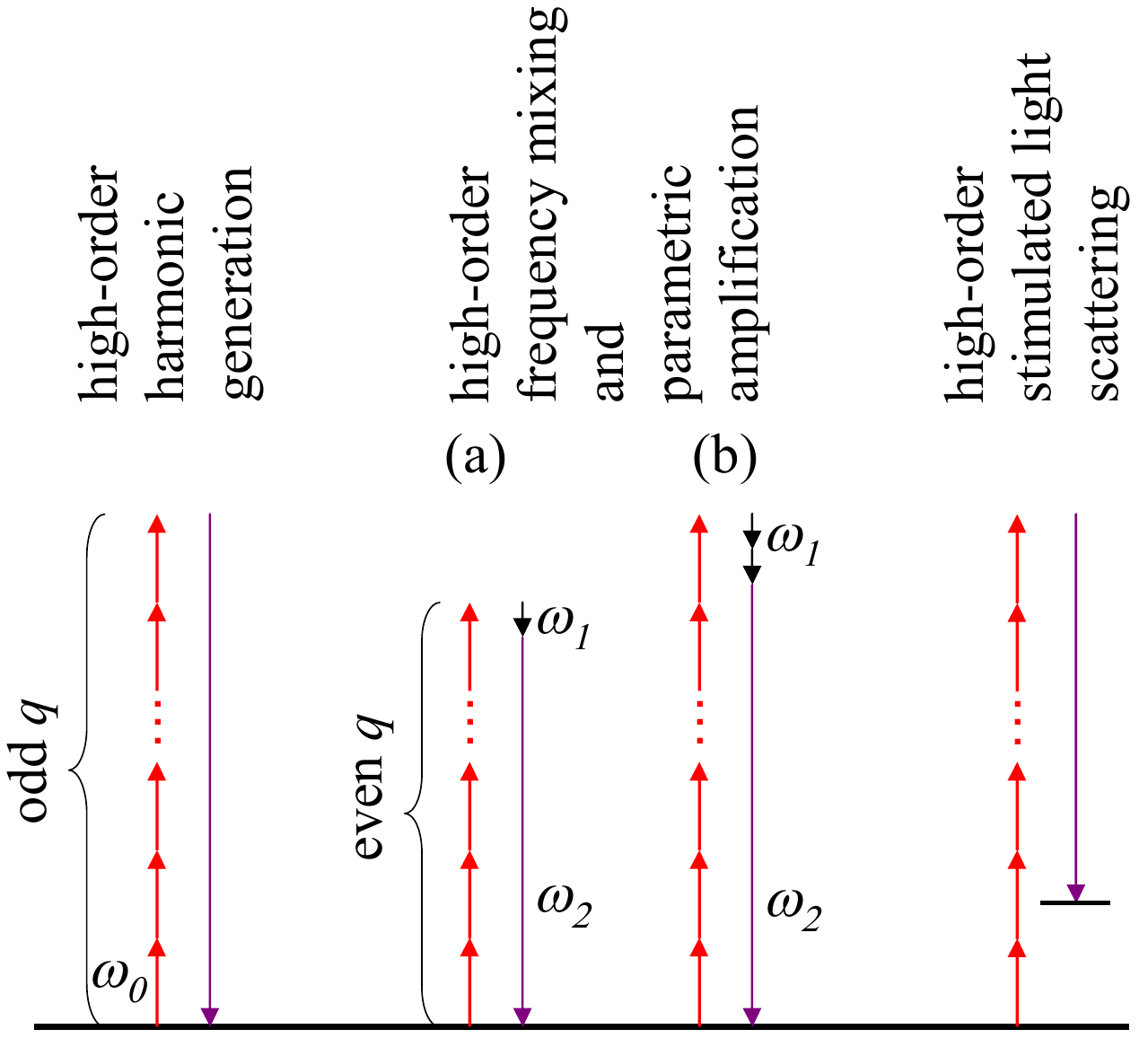}
\caption{(color online) 
The schematic of several high-order processes involving many photons from the pump field with frequency $\omega_0$ and few photons from the weaker fields with frequencies  $\omega_1$ and  $\omega_2$ .}
\label{Fig3}
\end{figure}

Below we study several high-order nonlinear processes illustrated with Figure~\ref{Fig3}. 

{ \it Wave-mixing} is the process where $q$ photons from the intense field and $m$ photons from the weaker field are converted in one $\omega_2=q \omega_0+m \omega_1$ photon. 
From equation
~(\ref{P_fin}) one can see that the intensity of the generated field $\mathcal{E}_{2}$ is proportional to the $m$-th power of the intensity of the field $\mathcal{E}_{1}$. This explains the experimental findings of~\cite{Worner_non-collinear} where for XUV generation this behavior was observed up to $m=6$. Another example is generation of THz radiation using a frequency-tunable half harmonic of a femtosecond pulse~\cite{Vvedenskii}. This process can be understood as difference frequency generation (one fundamental photon plus $m=-2$ half harmonic photons). The experimentally found dependence of the THz yield on 
 the weak half harmonic intensity is quadratic, as it should be for the process with $|m|=2$. 

From equation
~(\ref{P_fin}) one can see that the detuning from the phase-matching for a process of mixing of $q$ photons from one beam and $m$ photons from the other is, irrespective of the intensity ratio:
$\Delta \mathbf{k}_{q,m}=q \mathbf{k}_{0}+ m\mathbf{k}_{1} - \mathbf{k}_{2}$
 where $k_2$ is the wave-vector at the frequency $\omega_2$.
In~\cite{Platonenko_non-collinear}
 the phase-matching for the non-collinear HHG ($\omega_1=\omega_0$, $\mathbf{k}_{1} \ne \mathbf{k}_{0}$) was considered. Here we assume $\omega_1 \ne \omega_0$, co-directed  $\mathbf{k}_{1}$ and  $\mathbf{k}_{0}$,  plasma and/or capillary contribution to the dispersion dominates,
the plasma frequency $\omega_{pl}<<\omega_0$,~$\omega_1$. Then
$
\Delta k_{q,m}=\frac{\omega_{pl}^2}{2c}
\left( -\frac{q}{\omega_0}-\frac{m}{\omega_1}+\frac{1}{q \omega_0+m \omega_1}\right)
$. 
 Omitting the last term we find that for $m=-\frac{\omega_1}{\omega_2}$ the detuning is zero {\it irrespective of the plasma frequency}, thus irrespective of its density. This is very important because the medium ionization always accompanies processes in strong laser field and thus the plasma density is time-varying. 

The possibility of the phase-matching optimization in the difference-frequency mixing in plasma, involving few photons from the two waves was first shown in~\cite{Shkolnikov, Milchberg} and further studied in~\cite{Kapteyn_2007}. For the case $\omega_0=2\omega_1$,  $q=6$, $m=-3$ the phase-matching was experimentally demonstrated in~\cite{Chichkov_plsama}.
However, the case of $\omega_0=2\omega_1$ is hardly perspective for solving phase-matching problem for really high harmonics because along with the optimal ($m=-q/2$) polarization wave, many other polarization waves are generated. This is not the case when only few waves with different (small) $m$ are generated. For $|m|<<|q|$ it should be $\omega_1<<\omega_0$, that is why in Fig.~\ref{Fig1} and~\ref{Fig2} we present the numerical results for this case. 
To achieve phase-matched generation of XUV using given number $q$ and lowest $|m|$, for the case of even $q$ one should use frequency $\omega_1=\omega_0/q$ and  $m=-1$ process, and for odd $q$ one should use frequency $\omega_1=2 \omega_0/q$ and $m=-2$ process, see Fig.\ref{Fig3}. 

{\it Parametric amplification and generation.}
The process of the parametric amplification is described with the same equations as the difference-frequency generation, the difference is in the input conditions: the former assumes that initially there is one intense pump field and both generated fields are initially weak, while the latter assumes two initial intense fields~\cite{Shen}. 

{\it (i) generation of two photons.}
Let us consider the process in which $q$ photons from the initial field are converted into one $\omega_1$ photon and one $\omega_2$ photon, see Fig.~\ref{Fig3}a. In this case $q$ is an even number (note that for $q=2$ this is a well-known process of four-wave mixing; however, for higher $q$ the analogy with the four-wave mixing is hardly helpful).  For $m=-1$ we have from equation~(\ref{P_fin}):
\begin{equation}
\begin{array}{l}
P^{NL}(\omega_{1,2},z)=\kappa_q^{(1)}(\omega_{1,2}=q \omega_0-\omega_{2,1}) \times \\
\exp\{iq(k_0 z-\omega_0 t+ \varphi_0)\} E_{2,1}^{*}
\end{array}
\label{P_par}
\end{equation} 
Substituting this polarization into the propagation equation we have:
\begin{equation}
\partial \mathcal{E}_{1,2}/\partial z=i 2 \pi k_{1,2} \kappa_q^{(1)}(\omega_{1,2})  \mathcal{E}_{2,1}^{*} 
\exp\{i \Delta k z+i q \varphi_0\}
\label{propagation_equation2}
\end{equation}
where 
in the considered case is 
$\Delta k=q k_0 - k_1 -k_2$.

These equations are similar to those describing the parametric amplification in the perturbative nonlinear optics (see~\cite{Shen}, part 9.1).  However, an important difference from this case is that the susceptibility $\kappa_q^{(m)}$ is complex, and its phase is not negligible. For instance, the phase of $\kappa_q^{(0)}$ is well-understood within the recollision picture (see~\cite{Salieres} and references therein). The details of the solution of the propagation equations~(\ref{propagation_equation2}) are presented in Appendix III. We obtain the solution in which the exponential growth 
dominates after a certain propagation distance.
\begin{figure}  
\centering
\includegraphics [width=1.0\columnwidth] {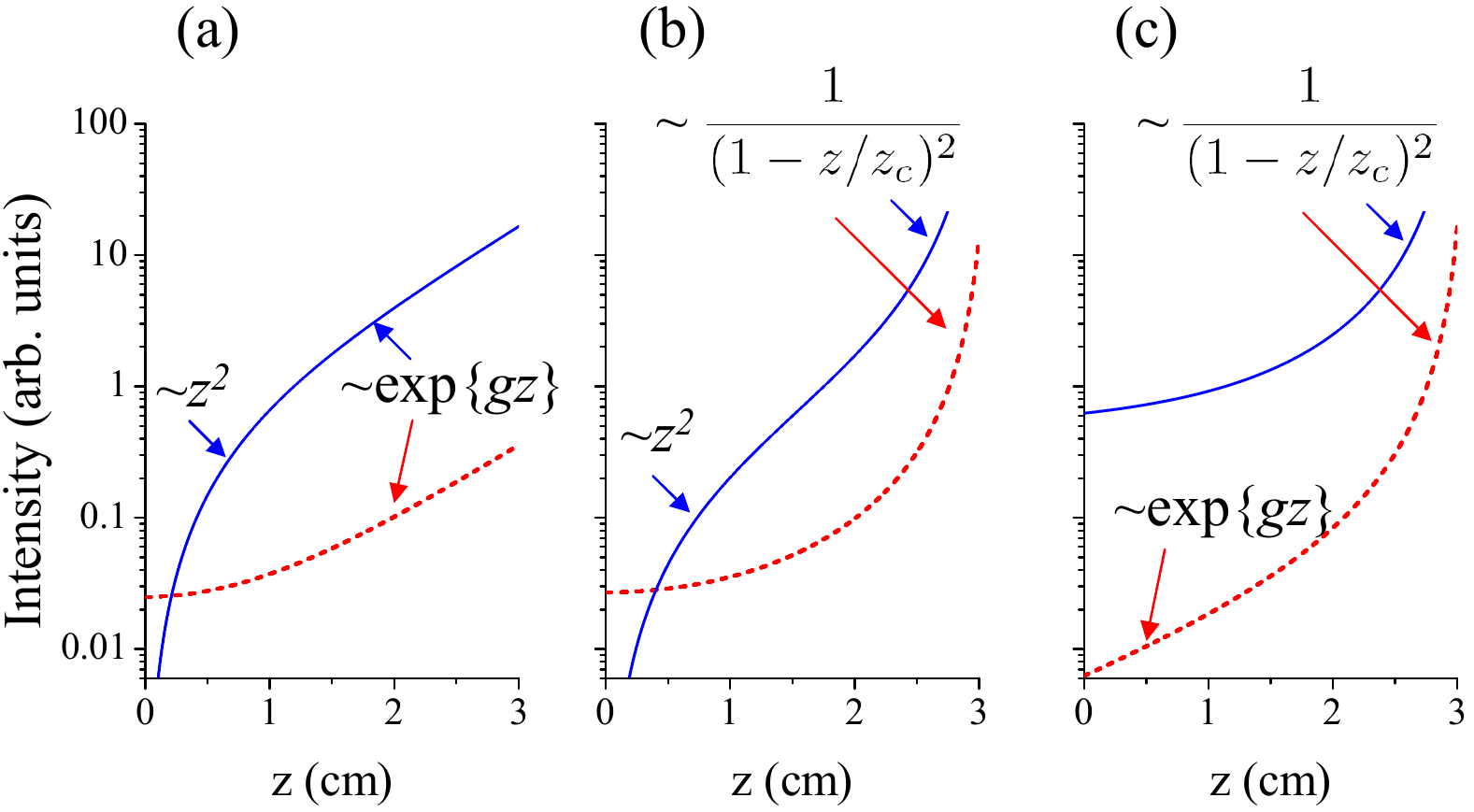}
\caption{(color online) 
Intensities of the fields $E_1$ (dotted red line) and $E_2$ (solid blue line) as a function of the propagation length for: two-photon parametric amplification (a), three-photon parametric amplification for different initial intensities of the fields (b and c).}
\label{Fig4}
\end{figure}

Our numerical calculations show that within the plateau region $|\kappa_q^{(1)}|$ is almost independent on $q$. For the conditions of Fig.~\ref{Fig1} we find that for $q=12$  $|\kappa_q^{(1)}|=4.5\times 10^{-26}n_{gas}$ CGS units, where $n_{gas}$ is the gas density (in cm$^{-3}$). According to equation~(\ref{Re_g}) (see Appendix III) this gives the exponential growth of the intensity with the increment $ g=1.35$~cm$^{-1}$ for the atmospheric pressure. Fig.~\ref{Fig4}(a) shows the fields' intensities calculated for these conditions, under zero incident intensity of the first field. One can see that initially ($z<1$~cm) the intensity of the first field growths quadratically, so the solution describes difference frequency generation, and for $z>2$~cm it grows exponentially. Note that the solution in the transition region can be less smooth depending on the initial phases of the fields and phases of the induced susceptibilities.  

The found values of increment $g$ show that the high-order parametric process can hardly be observed in atomic gas jets, but it can be observed in larger targets (capillaries and cells) especially under high gas pressure (pressures up to tens of atmospheres were used in recent experiments~\cite{Baltushka}). Moreover, this process can be very efficient in parametric generator. Such parametric generator can be only singly-resonant one because of extremely low efficiency of optical elements in XUV.
 Note that the intracavity HHG was demonstrated experimentally~\cite{intracavity_HHG} and now it  is actively used.


{\it (ii) generation of three photons.}
Another parametric process, which was not considered in the perturbative nonlinear optics, is the one where $q$ photons from the pump field are split into more than two photons. Here we shall study the process where these are two low-frequency photons ($\omega_1$) and one XUV photon ($\omega_2$). Here $q$ should be odd, so under $\omega_1<<\omega_0$ the generated XUV frequency $\omega_2$ is close to the q-th harmonic frequency. Thus, the naturally broadened HH line can provide an effective seeding for this process, and this determines its practical importance. In Appendix IV we obtain the following propagation equations describing this process:
$$
\begin{array}{l}
\partial \mathcal{E}_{1}/ \partial z=  A_1 \mathcal{E}_{1}^{*} \mathcal{E}_{2}^{*} \exp\{i \Delta k z+i q \varphi_0 \} \\
\partial \mathcal{E}_{2}/\partial z=A_2  \left(\mathcal{E}_{1}^{*} \right)^2 \exp\{i \Delta k z+i q \varphi_0 \}
\end{array}
$$
where  $\Delta k=q k_0 - 2 k_1 -k_2$. We solve it analytically for $\Delta k=0$. The main property of the found analytical solution is the {\it hyperbolic} growth after a certain propagation distance: $\mathcal{E}_{1,2}(z)\propto 1/(1-z/z_c)$. Certainly, when $z$ is close to $z_c$ the approximation of the given pump field fails, but the efficiency of the process should be high for such $z$.   

Fig.~\ref{Fig4} shows the fields' intensities calculated for zero incident first field (b) and small (but not zero) incident seeding second field (c); in contrast to the generation of two photons, here the propagation equations are not symmetric with respect to the fields. In panel (b) one can see that initially the intensity of the first field grows quadratically, so the solution describes difference frequency generation. For the case presented in panel (c) initially the second field grows exponentially from the seeding. The following hyperbolic growth of the intensities is common for both cases. Again, the solutions in the transition region can be less smooth depending on the initial phases of the fields and phases of the induced susceptibilities.  

{\it Stimulated scattering.}
The schematic of the Stokes wave generation due to the high-order stimulated scattering is presented in the figure~\ref{Fig3}. Similar to the perturbative case (see~\cite{Shen}, paragraph 10.3 ),  nonlinear polarization can be written as:
$
P^{NL}(\omega_{1})=\kappa_{R}^{(1)}(\omega_1) E_{1} (\omega_{1})
$
where $\kappa_{R}^{(1)}(\omega_1)$ is the induced Raman susceptibility. Note that here we do not specify the number of laser quanta $q$ involved in the process. Practically $\kappa_{R}^{(1)}$ can be calculated exactly in the same way as it was done above, namely as the limit of $P(\omega_1, \mathcal{E}_1)/\mathcal{E}_1$ for weak $\mathcal{E}_1$. The propagation equation for the Stokes wave is:
$
\left( \frac{\partial}{\partial z}+ \alpha\right)  \mathcal{E}_1=i 2 \pi  k_1 \kappa_{R}^{(1)}(\omega_1)   \mathcal{E}_{1}
$ 
 where $\alpha$ is linear absorption coefficient for frequency $\omega_1$. Its solution is $\mathcal{E}_1=\mathcal{E}_1(0) \exp\{g_R z - \alpha z\}$ where $ g_R=i 2 \pi  k_1 \kappa_{R}^{(1)}(\omega_1) $. 
The main practical advantage of this process for effective XUV generation is the absence of the phase-matching problem. Our approach can be useful for describing XUV amplification found in numerical studies~\cite{Serrat_2013, Serrat_2015}. 




In conclusion, in this paper we suggest the formalism of the nonperturbatively induced susceptibilities which allows writing the propagation equations in a form similar (but not identical) to the one used in perturbative nonlinear optics. Under some limitations we derive the analytical solutions for the propagation equations, describing several high-order optical properties. In particular, the super-exponential (hyperbolic) growing solutions are found for the three-photon parametric amplification, never studied earlier to the best of our knowledge. The numerically found susceptibilities are too low to make these processes observable in small dilute targets like gas jets, but in larger targets with higher densities (high-pressure gas filled capillaries) such processes are feasible, opening a way of efficient generation of coherent XUV. Finally, we discuss published numerical and experimental results where some processes from the nonperturbative nonlinear optics might have already been observed.

This study was supported by the "Extreme fields and applications" program of the Presidium of RAS.

\clearpage
\onecolumngrid 
\section*{Appendix I. Theory}

For completeness in this Appendix we shall start deriving equation~(\ref{P_fin}) from the very beginning. The microscopic polarization $P(t)$ is a {\it functional} of the external field $E=E_0+E_1$

$$
P(t)=\Phi[E]=\Phi[E_0+E_1]
$$

The functional can be expanded in the Taylor series~\cite{Volterra} (for some physical applications of such expansion, see ~\cite{Feynman,Engel}):

$$
P(t)=P^{(0)}(t)+P^{(1)}(t)+P^{(2)}(t)+ \cdots
$$

where

\begin{equation}
P^{(0)}(t)=\Phi[E_0],
\label{P0_a}
\end{equation}

\begin{equation}
P^{(1)}(t)=\int_0^{+\infty} d \tau \left. \frac{\delta \Phi}{\delta E} \right|_{E=E_0(t-\tau)} E_1(t-\tau)  +c.c.
\label{P1_a}
\end{equation}

\begin{equation}
P^{(2)}(t)= \int_0^{+\infty} d \tau \int_0^{+\infty} d \tau' \left. \frac{1}{2}\frac{\delta^2 \Phi}{\delta E^2} \right|_{E=E_0(t-\tau')} E_1(t-\tau) E_1(t-\tau')  +c.c.
\label{P2_a}
\end{equation}

here $\frac{\delta \Phi}{\delta E}$, $\frac{\delta^2 \Phi}{\delta E^2} $ are the {\it functional} derivatives of the functional $\Phi$ over the function $E$.

Let us consider fields
$$
E_{0,1}=\mathcal{E}_{0,1} \exp(-i \omega_{0,1} t)+ c.c.
$$

Periodicity of the field $E_0$ allows the expansion of the functional derivatives in equations~(\ref{P1_a}),~(\ref{P2_a}) in the Fourier series. 

\begin{equation}
\left. \frac{\delta \Phi}{\delta E} \right|_{E=E_0(t-\tau)}=\sum_q G^{(1)}_q(\mathcal{E}_0, \omega_0, \tau) \exp(-i q \omega_0 (t-\tau))
\end{equation}

\begin{equation}
\left. \frac{1}{2}\frac{\delta^2 \Phi}{\delta E^2} \right|_{E=E_0(t-\tau')}=\sum_q G^{(2)}_q(\mathcal{E}_0, \omega_0,\tau') \exp(-i q \omega_0 (t-\tau'))
\end{equation}

Substituting these expansions into equations~(\ref{P1_a}),~(\ref{P2_a}) we have:

\begin{equation}
\begin{array}{l}
P^{(1)}(t)=\sum_q \mathcal{E}_1 \exp(-i (q \omega_0  + \omega_1)t ) \kappa^{(+1)}_q + \mathcal{E}_1^* \exp(-i (q \omega_0  - \omega_1)t ) \kappa^{(-1)}_q  + c.c.
\end{array}
\label{P1_Fourier}
\end{equation}

\begin{equation}
P^{(2)}(t)=\sum_q \mathcal{E}_1^2 \exp(-i (q \omega_0  + 2 \omega_1)t ) \kappa^{(+2)}_q + \left (\mathcal{E}_1^*\right)^2 \exp(-i (q \omega_0 - 2 \omega_1)t ) \kappa^{(-2)}_q +  |\mathcal{E}_1|^2 \exp(-i q \omega_0  t ) \kappa^{(0,2)}_q + c.c.
\label{P2_Fourier}
\end{equation}

where

\begin{equation}
\kappa^{(\pm1)}_q (\mathcal{E}_0,\omega_0,\omega_1)=\int_0^{+\infty} d \tau G^{(1)}_q(\mathcal{E}_0, \omega_0, \tau) \exp(i q \omega_0 \tau\pm i \omega_1 \tau)
\label{kappa1}
\end{equation}

\begin{equation}
\kappa^{(\pm2)}_q (\mathcal{E}_0,\omega_0,\omega_1)=\int_0^{+\infty} d \tau \int_0^{+\infty} d \tau' G^{(2)}_q(\mathcal{E}_0, \omega_0,\tau') \exp(i q \omega_0 \tau' \pm i \omega_1 (\tau + \tau'))
\label{kappa2}
\end{equation}

\begin{equation}
\kappa^{(0,2)}_q (\mathcal{E}_0,\omega_0,\omega_1)=\int_0^{+\infty} d \tau \int_0^{+\infty} d \tau' G^{(2)}_q(\mathcal{E}_0, \omega_0,\tau') \exp(i q \omega_0 \tau') [\exp( i \omega_1 (\tau - \tau'))+\exp(- i \omega_1 (\tau - \tau'))]
\label{kappa3}
\end{equation}


Under certain conditions (see [M. L. Pons, R. Taieb, and A. Maquet {\it Phys. Rev. A} {\bf  54}, 3634 (1996)]  and references therein) the functional  $\Phi[E_0]$ in equation~(\ref{P0_a})  can be expanded in Fourier series according to the Floquet theorem: 
\begin{equation}
P^{(0)}(t)=\sum_q \kappa^{(0)}_q  (\mathcal{E}_0,\omega_0) \exp(-i q \omega_0 t )  + c.c.
\label{P0_Fourier}
\end{equation}

Thus the Fourier expansion of the microscopic response $P$ can be written as:

\begin{equation}
P(t)=\sum_{q=-\infty}^{+\infty} \sum_{m=-\infty}^{+\infty} \tilde \kappa_{q}^{(m)}  (\mathcal{E}_1)^{|m|}  \exp\{-iq\omega_0 t-im\omega_1 t\} + c.c.
\label{P_fin1}
\end{equation}

where

\begin{equation}
\tilde \kappa_{q}^{(m)} (\mathcal{E}_0,\mathcal{E}_1,\omega_0,\omega_1) = \sum_{j=0,2,\cdots}^{+\infty} \kappa_{q}^{(m,j)} (\mathcal{E}_0,\omega_0,\omega_1)|\mathcal{E}_1|^j  
\label{tilde_kappa_A}
\end{equation}
$$
\kappa_{q}^{(m,0)}\equiv \kappa_{q}^{(m)}
$$

From equation~(\ref{tilde_kappa_A}) one can see that terms with $j=2,4,...$ can be understood as those describing processes where one, two, etc. photons from the field $E_1$ were absorbed and the same number of photons of this field were emitted.

Now let us consider the fields 
\begin{equation}
E_{0,1}=\mathcal{E}_{0,1} \exp(-i \omega_{0,1} t+\mathbf{k}_{0,1}\mathbf{r}+\varphi_{0,1})+ c.c.
\end{equation}
with the real amplitudes $\mathcal{E}_{0,1}$.

Assuming the locality of the microscopic response and substituting $-\omega_{0,1} t \rightarrow -\omega_{0,1} t +\mathbf{k}_{0,1}\mathbf{r}+\varphi_{0,1} $ in equation~(\ref{P_fin1}) we have

\begin{equation}
P(t)=\sum_{q=-\infty}^{+\infty} \sum_{m=-\infty}^{+\infty} \tilde \kappa_{q}^{(m)} (\mathcal{E}_1)^{|m|}  \exp\{iq(-\omega_0 t+\mathbf{k}_{0}\mathbf{r}+\varphi_0)+im(-\omega_1 t+\mathbf{k}_{1}\mathbf{r}+\varphi_1) \}  + c.c.
\end{equation}

Thus we have derived equation~(\ref{P_fin}).

\section*{Appendix II. Some properties of the induced susceptibilities}

In this paper we consider only the centrally symmetric case (the medium is centrally symmetric and the fields are polarized in the same direction).

{\it (i)}  From the numerical TDSE solution we have found that while $q$ is in the plateau region $\kappa_q^{(m)}$ are comparable for different orders $q$.

{\it (ii)} For $\omega_1<<q \omega_0$ and in absence of resonances from equation~(\ref{kappa1})  one can see that 

\begin{equation}
\kappa_q^{(1)}(\omega_1) \approx \kappa_q^{(-1)}(\omega_1)
\end{equation}

From equation~(\ref{kappa2}) for $\kappa_q^{(2)}$ and from similar equations for higher-order induced susceptibilities we have under $m \omega_1<<q \omega_0$

\begin{equation}
\kappa_q^{(m)}(\omega_1) \approx \kappa_q^{(-m)}(\omega_1)
\end{equation}

For $m=1,2$ this can be seen from Fig.~\ref{Fig1}. Moreover, for higher $m$ the difference between  $\kappa_q^{(m)}$ and $\kappa_q^{(-m)}$ is visible. 

{\it (iii)} In the following studies we shall write explicitly the frequency of the response, using the following notations:
$$
\kappa^{(1)}(\omega=q \omega_0 \pm \omega_1) \equiv \kappa_q^{(\pm 1)}(\omega_1)
$$
$$
\kappa^{(2)}(\omega=q \omega_0 \pm 2 \omega_1) \equiv \kappa_q^{(\pm 2)}(\omega_1)
$$

The permutation symmetry of the second order nonlinear susceptibilities is~\cite{Shen}

\begin{equation}
\chi^{(2)*}(\omega=\omega_1+\omega_2)=\chi^{(2)}(\omega_1=-\omega_2+\omega)=\chi^{(2)}(\omega_2=\omega-\omega_1)
\end{equation}

In the approximation of the given pump field the first equation does not have its analog for the induced susceptibilities. However, the second equation has the analog:

 \begin{equation}
|\kappa^{(1)}(\omega_2=q \omega_0-\omega_1)|=|\kappa^{(1)}(\omega_1=-\omega_2+q \omega_0)|
\label{permutation_symmetry}
\end{equation}

Note that the equation takes place only for the absolute values. We shall derive this equation in Appendix III. It can be also derived directly from the Manley-Rowe relations as it is done in~[L.D. Landau, E.M. Lifshitz, Electrodynamics of Continuous Media (Volume 8 of A Course of Theoretical Physics), Pergamon Press, 1960, \S108] for $\chi$.



\section*{Appendix III. Parametric generation of two photons.}

Let us consider the process in which $q$ photons from the initial field are converted into one $\omega_1$ photon and one $\omega_2$ photon, see Fig.~\ref{Fig3}a. In this case $q$ is an even number (note that for $q=2$ this is a well-known process of four-wave mixing; however, for higher $q$ the analogy with the four-wave mixing is hardly helpful). In the plane-wave approximation and slowly-varying amplitude approximation
$
E_{1,2}=\mathcal{E}_{1,2}(z)\exp\{i(k_{1,2}z-\omega_{1,2}t )\}
$
and the complex amplitudes of the fields are described with the following propagation equations:
\begin{equation}
\frac{\partial \mathcal{E}_{1,2}}{\partial z}=\frac{i 2 \pi \omega_{1,2}^2}{k_{1,2}c^2} P^{NL}( \omega_{1,2},z) \exp\{-i(k_{1,2}z-\omega_{1,2}t\}
\label{propagation_equation1}
\end{equation}
where $P^{NL}$ is the nonlinear part of the polarization. For $m=-1$ we have from equation~(\ref{P_fin}):
\begin{equation}
P^{NL}(\omega_{1,2},z)=\kappa_q^{(1)}(\omega_{1,2}=q \omega_0-\omega_{2,1}) 
\exp\{iq(k_0 z-\omega_0 t+ \varphi_0)\} E_{2,1}^{*}
\label{P_par_A}
\end{equation} 
Substituting this equation into equation~(\ref{propagation_equation1}) and assuming in the denominator $k_{1,2}\approx \omega_{1,2}/c $ we have:
\begin{equation}
\frac{\partial \mathcal{E}_{1,2}}{\partial z}=i 2 \pi k_{1,2} \kappa_q^{(1)}(\omega_{1,2})  \mathcal{E}_{2,1}^{*} 
\exp\{i \Delta k z+i q \varphi_0\}
\label{propagation_equation2_A}
\end{equation}
where $\Delta k$ 
in the considered case is 
$\Delta k=q k_0 - k_1 -k_2$.

Propagation equations~(\ref{propagation_equation2_A}) are similar to those describing the parametric amplification in the usual (perturbative) nonlinear optics (see~\cite{Shen}, part 9.1). Note that this similarity can also be seen from the fact that the perturbative nonlinear response in the presence of the pump field $E_3$ with the frequency $\omega_3=\omega_1+\omega_2$: $P^{NL,per} (\omega_{1,2})=\chi^{(2)}E_{2,1}^{*}E_3$ can be converted in equation~(\ref{P_par_A}) substituting $\chi^{(2)}E_3 \rightarrow \kappa_q^{(1)}\exp\{iq(k_0  z-\omega_0 t+\varphi_0)\}$. However, an important difference from the perturbative case is that the susceptibility $\kappa_q^{(m)}$ is complex, and its phase is not negligible. For instance, the phase of $\kappa_q^{(0)}$ is well-understood within the recollision picture (see~\cite{Salieres} and references therein).

 Let us introduce:

\begin{equation}
A_{1,2}=i 2 \pi k_{1,2} \kappa_q^{(1)}(\omega_{1,2}) 
\label{A}
\end{equation}

Equations~(\ref{propagation_equation2_A}) are rewritten as:

\begin{equation}
\frac{\partial \mathcal{E}_{1,2}}{\partial z}=A_{1,2} \mathcal{E}_{2,1}^{*} 
\exp\{i \Delta k z+i q \varphi_0 \}
\label{propagation_equation3}
\end{equation}

Below we shall consider the case of the exact phase-matching $\Delta k=0$. Let us introduce:

$$
\mathcal{E}_{1,2}=u_{1,2} \exp\{i \varphi_{1,2}\}
$$

\begin{equation}
A_{1,2}=a_{1,2} \exp\{i \theta_{1,2}\}
\label{Aa}
\end{equation}

where $u_{1,2}$, $\varphi_{1,2}$, $a_{1,2}$, $\theta_{1,2}$ are real. Introducing

$$
\psi=\theta_1+q \varphi_0-\varphi_1-\varphi_2 
$$

$$
\Delta \theta= \theta_2- \theta_1
$$
we have from equations~(\ref{propagation_equation3}):

\begin{equation}
\begin{array}{l}
\frac{\partial u_1}{\partial z}=a_1 u_2 \cos(\psi) \\
\frac{\partial u_2}{\partial z}=a_2 u_1 \cos(\psi+\Delta \theta) \\
u_1 \frac{\partial \varphi_1}{\partial z}=a_1 u_2 \sin(\psi)\\
u_2 \frac{\partial \varphi_2}{\partial z}=a_2 u_1 \sin(\psi+\Delta \theta)
\end{array}
\label{propagation_equation4}
\end{equation}

In the absence of resonances the number of quanta generated at frequencies $\omega_1$ and $\omega_2$ should be equal:

$$
\frac{1}{\omega_1}\frac{\partial (u_1^2)}{\partial z}=\frac{1}{\omega_2}\frac{\partial (u_2^2)}{\partial z}
$$

From this equation and the first pair of equations~(\ref{propagation_equation4}) we find:

\begin{equation}  
\label{cos_psi}
\frac{a_1}{\omega_1}\cos(\psi) = \frac{a_2}{\omega_2}\cos(\psi+ \Delta \theta)
\end{equation}

Under $\Delta \theta = 0$ this gives 

\begin{equation} 
\frac{a_1}{\omega_1}=\frac{a_2}{\omega_2}
\label{a_w}
\end{equation}

Under $\Delta \theta \ne 0$ equation~(\ref{cos_psi}) can be satisfied only under $\psi=const$ (for every solution of the system of linear equations~(\ref{propagation_equation4}); for different solutions $\psi$ can be different). Thus $\frac{\partial \psi}{\partial z}=-\frac{\partial \varphi_1}{\partial z}-\frac{\partial \varphi_2}{\partial z}=0$ and from the second pair of equations~(\ref{propagation_equation4}) we find:

\begin{equation} 
\frac{a_1}{\omega_1}=\frac{a_2}{\omega_2}
\label{a_w_1}
\end{equation}

This, this equation is valid irrespectively on $\Delta \theta$. From equations~(\ref{A}),~(\ref{Aa}),~(\ref{a_w_1}) we have  

\begin{equation}  
\label{permutation_symmetry_corr}
|\kappa_q^{(1)}(\omega_{1})|=|\kappa_q^{(1)}(\omega_{2})|
\end{equation}

This equation presents the permutation symmetry of induced susceptibilities of the first order. Let us denote $|\kappa_q^{(1)}|=|\kappa_q^{(1)}(\omega_{1,2})|$, so $a_{1,2}=2 \pi k_{1,2} |\kappa_q^{(1)}|$.

The solution of equation~(\ref{propagation_equation3}) is:

$$
\mathcal{E}_{1,2}(z)=\mathcal{E}_{1,2}^{+}(z)+\mathcal{E}_{1,2}^{-}(z)
$$
$$
\mathcal{E}_{1}^{\pm}(z)=\mathcal{E}_{1}^{\pm}(0) \exp\{g^{\pm}z + i \varphi_1^{\pm}(0)\}
$$
$$
\mathcal{E}_{2}^{\pm}(z)=\mathcal{E}_{2}^{\pm}(0) \exp\{\left(g^{\pm}\right)^{*}z + i \varphi_2^{\pm}(0)\}
$$
$$
\mathcal{E}_{1}^{\pm}(0)=\left| \frac{\alpha \mathcal{E}_{2}(0)+ \mathcal{E}_{1}^{*}(0)\exp\{i(\theta_1+q \varphi_0-\psi^{\pm})\}}{2}\right|
$$
$$
\mathcal{E}_{2}^{\pm}(0)=\mathcal{E}_{1}^{\pm}(0)/\alpha
$$
$$
\varphi_2^{\pm}(0)=\arg\left(  \frac{\alpha \mathcal{E}_{2}(0)+ \mathcal{E}_{1}^{*}(0)\exp\{i(\theta_1+q \varphi_0-\psi^{\pm})\}}{2}\right)
$$
$$
\varphi_1^{\pm}(0)=\theta_1+q \varphi_0-\psi^{\pm}-\varphi_2^{\pm}(0)
$$
where $g^{\pm}=2 \pi \sqrt{k_1 k_2} |\kappa_q^{(1)}| \exp{i \psi^{\pm}}$, $ \alpha=\sqrt{\omega_1/\omega_2}$, $\psi^+=-\Delta \theta/2$, $\psi^-=-\Delta \theta/2+\pi$

For the phase-matched process with $\omega_1=\omega_0/q$ we have 
\begin{equation}
| Re (g^{\pm})|=2 \pi k_0 |\kappa_q^{(1)}|  |\cos(\Delta \theta/2)| 
\label{Re_g}
\end{equation}

\section*{Appendix IV. Parametric generation of three photons.}

The nonlinear polarizations for this process are:

\begin{equation}
P^{NL}(\omega_{1})=\kappa_q^{(2)}(\omega_1=q\omega_0-\omega_1-\omega_2)\exp\{iq(k_0 z-\omega_0 t+ \varphi_0)\} E_{1}^{*} E_{2}^{*}
\label{P_par_1}
\end{equation}

\begin{equation}
P^{NL}(\omega_{2})=\kappa_q^{(2)}(\omega_2=q\omega_0-2 \omega_1)\exp\{iq(k_0 z-\omega_0 t+ \varphi_0)\} \left( E_{1}^{*} \right)^2 
\label{P_par_2}
\end{equation}

Let us denote:

\begin{equation}
A_1=i 2 \pi k_{1} \kappa_q^{(2)}(\omega_1=q\omega_0-\omega_1-\omega_2)
\label{A1}
\end{equation}

\begin{equation}
A_2=i 2 \pi k_{2} \kappa_q^{(2)}(\omega_2=q\omega_0-2\omega_1) 
\label{A2}
\end{equation}

 The propagation equations are written as:

\begin{equation}
\frac{\partial \mathcal{E}_{1}}{\partial z}=  A_1 \mathcal{E}_{1}^{*} \mathcal{E}_{2}^{*}
\exp\{i \Delta k z+i q \varphi_0 \}
\label{propagation_equation3_1}
\end{equation}

\begin{equation}
\frac{\partial \mathcal{E}_{2}}{\partial z}=A_2  \left(\mathcal{E}_{1}^{*} \right)^2
\exp\{i \Delta k z+i q \varphi_0 \}
\label{propagation_equation3_2}
\end{equation}

where  $\Delta k=q k_0 - 2 k_1 -k_2$. Below we shall consider the case of the exact phase-matching 
$\Delta k =0$. 

Let us introduce:

$$
\mathcal{E}_{1,2}=u_{1,2} \exp\{i \varphi_{1,2}\}
$$

\begin{equation}
A_{1,2}=a_{1,2} \exp\{i \theta_{1,2}\}
\label{Aa_three-ph}
\end{equation}

where $u_{1,2}$, $\varphi_{1,2}$, $a_{1,2}$, $\theta_{1,2}$ are real. Introducing

$$
\psi=\theta_1+q \varphi_0-2\varphi_1-\varphi_2 
$$

$$
\Delta \theta= \theta_2- \theta_1
$$
we have from equations~(\ref{propagation_equation3_1}),~(\ref{propagation_equation3_2}):

\begin{equation}
\begin{array}{l}
\frac{\partial u_1}{\partial z}=a_1 u_1 u_2 \cos(\psi) \\
\frac{\partial u_2}{\partial z}=a_2 u_1^2 \cos(\psi+\Delta \theta) \\
u_1 \frac{\partial \varphi_1}{\partial z}=a_1 u_1 u_2 \sin(\psi)\\
u_2 \frac{\partial \varphi_2}{\partial z}=a_2 u_1^2 \sin(\psi+\Delta \theta)
\end{array}
\label{propagation_equation4_three-ph}
\end{equation}

In the absence of resonances the number of quanta generated at the frequency $\omega_1$ should be twice that of quanta at the frequency $\omega_2$:

$$
\frac{1}{\omega_1}\frac{\partial (u_1^2)}{\partial z}=\frac{2}{\omega_2}\frac{\partial (u_2^2)}{\partial z}
$$

From this equation and the first pair of equations~(\ref{propagation_equation4_three-ph}) we find:

\begin{equation}  
\label{cos_psi_three-ph}
\frac{a_1}{\omega_1}\cos(\psi) = 2 \frac{ a_2}{\omega_2}\cos(\psi+ \Delta \theta)
\end{equation}

{\it Case 1: $\Delta \theta \ne 0$.} Equation~(\ref{cos_psi_three-ph}) can be satisfied only under $\psi=const$:

\begin{equation}  
\label{tan_psi}
\tan(\psi) = \left[ \cos(\Delta \theta) -\frac{|\kappa_q^{(2)}(\omega_1)|}{2 |\kappa_q^{(2)}(\omega_2)|} \right] \frac{1}{\sin(\Delta \theta)}
\end{equation}

Thus 

\begin{equation}
\frac{\partial \psi}{\partial z} =-2 \frac{\partial \varphi_1}{\partial z}-\frac{\partial \varphi_2}{\partial z}=0
\label{psi_z}
\end{equation}

 and from this equation and the second pair of equations~(\ref{propagation_equation4_three-ph}) we find:

\begin{equation}
u_1(z)/u_2(z)=const
\label{u_1_u_2}
\end{equation}

The solution of the first pair of equations~(\ref{propagation_equation4_three-ph}) satisfying~(\ref{u_1_u_2})  is: 

\begin{equation}
u_{1,2}(z)=u_{1,2}(0) \frac{1}{1-z/z_c}
\label{catastrophe}
\end{equation}
where
\begin{equation}
z_c=\frac{1}{u_2(0) a_1 \cos(\psi)}
\label{z_catastrophe}
\end{equation}
and
$$
\frac{u_1(z)}{u_2(z)}=\sqrt{\frac{a_1 \cos(\psi)}{a_2 \cos(\psi+ \Delta \theta)}}
$$

Taking into account equation~(\ref{cos_psi_three-ph}), this equation can be rewritten as

\begin{equation}
\frac{u_1(z)}{u_2(z)}=\sqrt{\frac{2 \omega_1}{\omega_2}}
\label{u_1_u_2_res}
\end{equation}

Thus only if the initial conditions satisfy this equation, the solution satisfies equation~(\ref{u_1_u_2}). Finally, from equations~(\ref{psi_z}) ,~(\ref{u_1_u_2_res}), and the second pair of equations~(\ref{propagation_equation4_three-ph}) we find:

\begin{equation} 
\cos(\Delta \theta)= \frac{|\kappa_q^{(2)}(\omega_1)|}{| \kappa_q^{(2)}(\omega_2)|}-\frac{2 |\kappa_q^{(2)}(\omega_2)|}{ |\kappa_q^{(2)}(\omega_1)|}
\label{cos_Delta_Theta}
\end{equation}

Thus we find the connection between phases and absolute values of the second-order induced susceptibilities.

{\it Case 2: $\Delta \theta = 0$.} Equation~(\ref{cos_psi_three-ph}) is satisfied for every $\psi$; it is satisfied under:

\begin{equation}  
\label{cos_psi_three-ph1}
\frac{a_1}{\omega_1} = 2 \frac{ a_2}{\omega_2}
\end{equation}

So 

\begin{equation}  
\label{perm2}
|\kappa_q^{(2)}(\omega_1)|=2|\kappa_q^{(2)}(\omega_2)|
\end{equation}

This equation presents the permutation symmetry of the induced susceptibilities of the second order for $\Delta \theta = 0$. Note that equation~(\ref{cos_Delta_Theta}) turns to~(\ref{perm2}) when  $\Delta \theta = 0$.

Let us denote:

$$
\mathcal{E}_{1,2}(z)=\mathcal{E'}_{1,2} (z) \exp\{ i \varphi_{1,2}(0)\}
$$ 
so that $\mathcal{E'}_{1,2} (0)$ are real. Then equations~(\ref{propagation_equation3_1}),~(\ref{propagation_equation3_2}) can be written as:

\begin{equation}
\frac{\partial \mathcal{E'}_{1}}{\partial z}=a_1  \mathcal{E'}_{1}^{*} \mathcal{E'}_{2}^{*} \exp\{ i \psi(0)\}
\label{propagation_equation4_1}
\end{equation}

\begin{equation}
\frac{\partial \mathcal{E'}_{2}}{\partial z}=a_2  \left(\mathcal{E'}_{1}^{*} \right)^2 \exp\{ i \psi(0)\}
\label{propagation_equation4_2}
\end{equation}
where $\psi(0) =\theta_1+q \varphi_0-2 \varphi_1(0) - \varphi_2(0)$ (note that $\theta_1=\theta_2$ because we consider the case $\Delta \theta=0$). If initial phases $\varphi_1(0)$ and $\varphi_2(0)$ are chosen so that $\psi(0)=0$, the solution of equations~(\ref{propagation_equation4_1}),~(\ref{propagation_equation4_2}) remains real.  Substituting 

\begin{equation}
\mathcal{E'}_1(z)=\sqrt{\frac{2 \omega_1}{\omega_2}(\mathcal{E'}_{2}(z)^{2}- \tilde\mathcal{E}^2)}
\label{solution_E1}
\end{equation}

where $\tilde\mathcal{E}=\sqrt{\mathcal{E'}_{2}(0)^{2}-\frac{\omega_2}{2 \omega_1} \mathcal{E'}_{1}(0)^{2} }$, system of  equations~(\ref{propagation_equation4_1}),~(\ref{propagation_equation4_2}) is presented as a single equation (note also equation~(\ref{cos_psi_three-ph1})):

\begin{equation}
\frac{\partial \mathcal{E'}_{2}}{\partial z}=a_1(\mathcal{E'}_{2}^{2}- \tilde\mathcal{E}^2)
\label{propagation_equation5}
\end{equation}

Such equation is known as the Riccati equation. Its solution is: 

\begin{equation}
\mathcal{E'}_2(z)=\tilde\mathcal{E} \quad \frac{\mu \exp\{- \tilde g z\}+\exp\{ \tilde g z\}}{\mu \exp\{- \tilde g z\}-\exp\{ \tilde g z\}}
\label{solution_E2}
\end{equation}

where $\tilde g= a_1 \tilde\mathcal{E}$, $$\mu=\frac{\mathcal{E'}_2(0)+\tilde\mathcal{E}}{\mathcal{E'}_2(0)-\tilde\mathcal{E}}.$$ Thus equations ~(\ref{solution_E1}) and~(\ref{solution_E2}) give the solution. The typical feature of this solution is a hyperbolic growth after a certain propagation distance:
$$
\mathcal{E'}_{1,2}(z) \propto \frac{1}{1-z/z_{c}}
$$  

The solution goes to infinity at 
\begin{equation}
z_{c}=\frac{\ln(\mu)}{2 \tilde g}
\label{catastrophe_1}
\end{equation}

Thus we can see that in both cases $\Delta\theta \ne 0$ and $\Delta\theta = 0$ there is a hyperbolic growth of the solution after a certain propagation distance.

\end{document}